# Online Support System for Transnational Education

*Maria Spichkova, James Harland, and Ahmed Alharthi\**

School of Science, RMIT University, Australia
\*Presenting author, E-mail: {maria.spichkova,james.harland, ahmed.alharthi}@rmit.edu.au

**Abstract:** The number of students who travel abroad to study or are enrolled in a distance learning program outside their home country is growing. According to UNESCO, such students are called internationally mobile students (IMSs) and 5 destination countries accounted for almost 50% of IMSs: United States (18%), United Kingdom (11%), France (7%), Australia (6%), and Germany (5%). Internationalisation of the higher education has created the so-called borderless university, providing better opportunities for learning and increases the human and social sustainability.

In this paper, we propose an online system to support transnational education (TNE) in Australia. This system will consist of both an online module and quiz, and an automated assistance module. The first part will ensure that students understand TNE-related potential problems, such as the importance of submission deadlines, and provide students early hints and help to avoid the TNE-related problems. The second part will provide assistance with general issues related to studying and living in Australia, and more specific ones tailored to a specific institution. The ability to provide an immediate response when required will help overcome feelings of isolation, and provide appropriate advice on matters such as plagiarism and related issues, many of which arise from misunderstandings due to divergent cultural backgrounds and a lack of awareness about differences in socio-cultural environments.

## 1. Introduction

According to UNESCO statistics[1], Australia is the 4$^{th}$ largest provider of international education. In 2012, approx. 250,000 mobile students (6% of total mobile students) were hosted in Australian universities. Internationalisation of the higher education has created the so-called 'borderless university' (Sadiki, 2001; Knight, 2015), which provides better opportunities for learning and increases human and social sustainability (Woodcraft, 2012; Vallance et al., 2011, Penzenstadler et al., 2012). This also means new challenges due to the diversity of cultural and legal requirements. One obstacle to successful transnational teaching and learning is the diversity of cultural, technical and educational backgrounds of both academics and students. The goal of our approach is to help students and academics to overcome this obstacle, and to possibly convert this obstacle into strength. Naturally the first step in developing a potential solution to this problem is to understand it in appropriate detail.

There are a number of case studies and analytical works on the potential and real issues within transnational education. For example, K. Gribble and C. Ziguras (2003) pointed in their work to general transnational education issues that teaching staff might have and suggested additional training for teachers. I. Dunn and M. Wallace (2006) presented a case study on the preparedness and experiences of the Australian academics and transnational teaching. Another study has shown that transnational teaching has the capacity to transform educators, especially if they particularly recognise the impact of cultural differences in the teaching and learning process (Hoare, 2013).

---

[1] http://www.uis.unesco.org/Education/Pages/international-student-flow-viz.aspx

Most of the case studies and other works focus on teaching aspect of education. However, the education process consists of two highly cooperating parts: *teaching and learning*. Moreover, the number of students involved in the process is much higher than the number of teachers. For this reason we suggest studying the problem of diversity from the students' perspective, and hence to analyse potential solutions with this in mind.

RMIT University, like many Australian universities, currently provides a number of orientation events for new students, as well as a webpage with comprehensive information on the study at the university[2]. However, both options are passive experiences for the students, and there is no examination of whether the students are really familiar with the corresponding material. The current webpage also provides only a small amount of information related to the transnational aspects of education. For instance, the RMIT myCommunity is providing a few topics such as Tech Support and New Academic Street project discussion having a large number of unsolved posts and/or with no replies. Moreover, students with limited knowledge of English might find it difficult to follow the orientation session and to ask questions if something is unclear to them. In our recent work (Alharthi et al., 2015, Spichkova and Schmidt, 2015) we have presented an approach to requirements specification and analysis for eLearning systems and for geographically distributed software systems. Systems for eLearning are usually developed for use within different organisations or even different countries. It is not at all unusual to find that the specific requirements differ between organisations or countries for technical, cultural, or legal reasons. The challenge is to deal with this diversity in a systematic way, avoiding contradictions and non-compliance.

In our current work, we intend to extend our approach to systems requirements engineering to the development of a TNE support system, consisting of (1) an online module and quiz, and (2) an automated assistance module, cf. Figure 1. The system will provide an immediate response when required and appropriate advice on matters such as plagiarism and related issues, many of which arise from misunderstandings due to divergent cultural backgrounds and a lack of awareness about differences in socio-cultural environments.

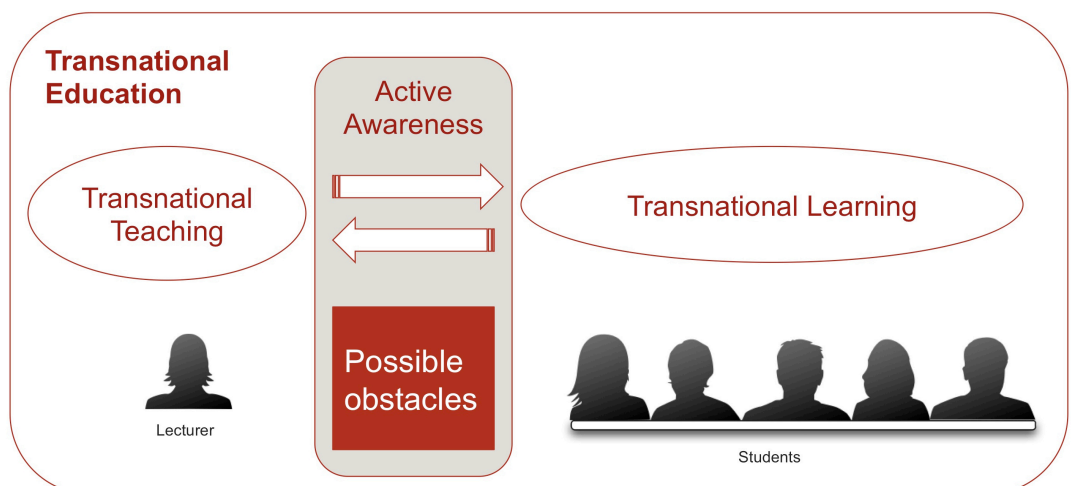

Figure 1. Transnational education: Transnational teaching + transnational learning

---

[2] https://www.rmit.edu.au

## 2. TNE Online Support system

The general architecture of the proposed TNE online support system is in Figure 2. This system will consist of two core parts:

- An online module on TNE related topics with a corresponding quiz designed to ensure that students understand TNE-related potential problems, such as the importance of submission deadlines, and provide students early hints and help to avoid the TNE-related problems.
- An automated assistance module to provide assistance with general issues related to studying and living in Australia, and more specific ones tailored to a specific institution.

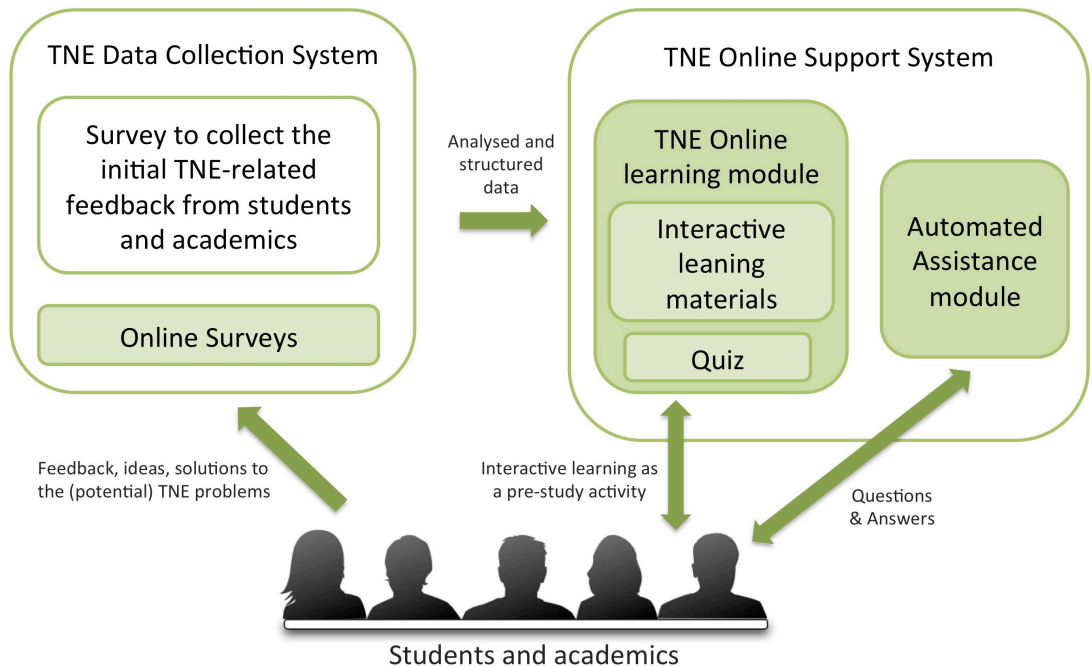

Figure 2. Proposed TNE Online Support System

### 2.1 Collection and analysis of the TNE data

The first step is to collect TNE-related feedback from the students and academics, including not only the descriptions of current issues but also gathering from students some appropriate suggestions and solutions. This step should be inclusive for two reasons:

- Students might suggest some effective and creative solutions, which haven't been previously taken into account by academics,
- Including students in the design of the solution will increase their engagement with the university, which will generally have a positive impact on their studies.

Additionally, the system will provide an online survey available at any time during the semester as well as on holidays. This would help to collect related feedback, suggestions and solutions more efficiently.

### 2.2 TNE online support system

The second step is to develop the online support system. RMIT University already has a number of online modules which are used for the induction and training of new staff members, on topics such as Health and Safety. We propose a *TNE Online Module*, which should be completed by both staff and students at the beginning of their study, as a final stage of the enrollment process. If the corresponding quiz is compulsory (but has no influence on grades), the teaching staff can have much greater confidence that the students are really familiar with the corresponding TNE material.

This learning module would
- Help students to understand TNE-related potential problems.
- Provide hints and assistance with TNE-related issues.

This solution would be especially beneficial for the students without strong skills in English, as it does not rely on listening skills. Also, this module may assist not only students but also the teaching staff; they will have a summary of the pre-study activity indicating appropriate students' level. Then, teachers can reconfigure programs to accommodate these differences. For instance, students who may interact in the second language, and have different culture-bound affecting communication skills.

The *automated assistance module* would then provide advice to the students on questions related to studying or living in Australia. Students are often hesitant to ask questions in 'live' sessions, due to communication or language issues. Students (even local students) are also hesitant to ask questions in front of other people, either in person or on electronic forums such as Blackboard discussion boards or Facebook. Often this leads to the question not being asked at all, or via more private means such as email. This creates a new problem, in that a staff member needs to reply to each individual query. Given that such queries often arrive in bursts around the same time, it will usually take up to one or two days for the student to receive a reply. In contrast, our proposed system will be available at all times and provide immediate answers.

We stress that our proposed system is intended as a replacement for many of the typical queries asked by students, but not as a complete replacement for human advisors. There will always be some issues or problems that will require human intervention, but the expertise of those able to do so will be put to greatest effect if many repetitive and routine queries can be dealt with by a system of this kind.

The system might be also provided in different languages using automated translator, to reduce the gap of communication between students having different mother tongues, as well provide a sustainable solution for information storage, to avoid duplicating storage of similar questions submitted in different languages.

There are also a number of contexts at RMIT University in which such a system could be used. The most natural starting point is with onshore international students here in Melbourne. However, as RMIT has two campuses in Vietnam, as well as a number of programs taught offshore, this system may provide benefits for a wide variety of TNE students.

Rapid changes in society and technology are affecting education and economic (Stepanyan et al., 2013), so it is important is to ensure consistent processes for transnational education and improve cost-effectiveness of the transnational education. The proposed system would help onshore international students and teaching staff not only to sustain TNE market but also to improve educational attainment. By guaranteeing of sharing knowledge, engaging international and local students, and developing students and teaching staff could minimise diversity and culture shock. For instance, the acknowledgement of different learning styles could improve student satisfaction.

## 3. Conclusions

In this position paper, we have presented our vision of a TNE online support system. This system will consist of both an online module and quiz, and an automated assistance module. The first part will ensure that students understand TNE-related potential problems, such as the importance of submission deadlines, and provide assistance with TNE-related issues. The second part will provide assistance with general issues related to studying and living in Australia, and more specific ones tailored to a specific institution.

The ability to provide an immediate response when required will help overcome feelings of isolation, and provide appropriate advice on matters such as plagiarism and related issues, many of which arise from misunderstandings due to divergent cultural backgrounds and a lack of awareness about differences in socio-cultural environments. We believe that this system will help students and teachers to overcome potential obstacles, which come from the diversity of their backgrounds, especially from the cultural diversity.